\documentclass[10pt,conference]{IEEEtran}
\IEEEoverridecommandlockouts
\pdfoutput=1
\usepackage{mdframed}
\usepackage{subfigure}
\usepackage[ruled,linesnumbered]{algorithm2e}
\usepackage{xspace}
\usepackage{listings} 
\usepackage{multirow}
\usepackage{graphicx}
\usepackage{comment}
\usepackage{xcolor}
\usepackage{bbding}
\usepackage{tabularx,booktabs} % For formal tables
\usepackage{amsmath} 
\usepackage[utf8]{inputenc}
\usepackage{caption}
\usepackage{enumitem}
\usepackage{url}
\usepackage{array, environ}
\usepackage{cite}
\usepackage{amsmath,amssymb,amsfonts}
\usepackage{textcomp}
\usepackage{tikz}
\usetikzlibrary{shapes.geometric, arrows}
\tikzset{%
            procedure/.style = {rectangle, draw=black,
                           minimum width=2cm, minimum height=1cm,
                           text centered, align=left},
               action/.style = {procedure, fill=blue!30},
               fontscale/.style = {font=\relsize{#1}},
}
\tikzstyle{arrow} = [thick,->,>=stealth]

\def\FuDriver{\textit{IntelliGen}}
\def\EFL{\textit{Entry Function Locator}}
\def\FDS{\textit{Fuzz Driver Synthesizer}}
\def\FUDGE{\textit{FUDGE}}
\def\FUZZGEN{\textit{FuzzGen}}
\def\LIBFUZZER{\textit{LibFuzzer}}
\newcommand{\mysize}{}

\newcounter{obstacle}[obstacle]

\SetAlFnt{\small}
\SetAlCapFnt{\small}
\SetAlCapNameFnt{\small}
\setenumerate[1]{itemsep=0pt,partopsep=0pt,parsep=\parskip,topsep=5pt}

\setitemize[1]{itemsep=0pt,partopsep=0pt,parsep=\parskip,topsep=5pt}

\setdescription{itemsep=0pt,partopsep=0pt,parsep=\parskip,topsep=5pt}

\def\BibTeX{{\rm B\kern-.05em{\sc i\kern-.025em b}\kern-.08em
    T\kern-.1667em\lower.7ex\hbox{E}\kern-.125emX}}

\begin{document}

\title{
IntelliGen: Automatic Driver Synthesis for Fuzz Testing
}

\author{Mingrui Zhang$^{1}$, Jianzhong Liu$^{2}$, Fuchen Ma$^{1}$, Huafeng Zhang$^{3}$, Yu Jiang*\thanks{*Yu Jiang is the corresponding author. This research is sponsored in part by the NSFC Program (No. 62022046, U1911401, 61802223), National Key Research and Development Project (Grant No. 2019YFB1706200), the Huawei-Tsinghua Trustworthy Research Project (No. 20192000794).}$^{1}$\\KLISS, BNRist, School of Software, Tsinghua Universiy$^{1}$\\ShanghaiTech University$^{2}$\\Huawei Technologies Co. Ltd, Beijing, China$^{3}$}

\begin{comment}
\author{
\IEEEauthorblockN{1\textsuperscript{st} Mingrui Zhang}
\IEEEauthorblockA{\textit{School of Software} \\
\textit{Tsinghua University}\\
Beijing, China \\
\qquad zmr19@mails.tsinghua.edu.cn\qquad }
\and
\IEEEauthorblockN{2\textsuperscript{nd} Yuanliang Chen}
\IEEEauthorblockA{\textit{School of Software} \\
\textit{Tsinghua University}\\
Beijing, China \\
\qquad chenyuan17@mails.tsinghua.edu.cn\qquad }
\and
\IEEEauthorblockN{3\textsuperscript{rd} Fuchen Ma}
\IEEEauthorblockA{\textit{School of Software} \\
\textit{Tsinghua University}\\
Beijing, China \\
\qquad mafc19@mails.tsinghua.edu.cn\qquad }
\and
\IEEEauthorblockN{4\textsuperscript{th} Zhe Liu}
\IEEEauthorblockA{\textit{School of Software} \\
\textit{Tsinghua University}\\
Beijing, China \\
\qquad sduliuzhe@163.com\qquad }
\and
\IEEEauthorblockN{5\textsuperscript{th} Huafeng Zhang}
\IEEEauthorblockA{\textit{Software Assurance Group} \\
\textit{Huawei Technologies co. Ltd}\\
Beijing, China \\
\qquad ron.huafeng@gmail.com\qquad }
\and
\IEEEauthorblockN{6\textsuperscript{th} Yu Jiang}
\IEEEauthorblockA{\textit{School of Software} \\
\textit{Tsinghua University}\\
Beijing, China \\
\qquad jiangyu198964@126.com\qquad }
}
\end{comment}

\maketitle

% Use the following at camera-ready time to suppress page numbers.
% Comment it out when you first submit the paper for review.
%\thispagestyle{empty}

\begin{abstract}

% Fuzz很NB
Fuzzing is a technique widely used in vulnerability detection. The process usually involves writing effective fuzz driver programs, which, when done manually, can be extremely labor intensive. Previous attempts at automation leave much to be desired, in either degree of automation or quality of output.

% FuDriver是啥，工作流程，实验结果
In this paper, we propose \FuDriver{}, a framework that constructs valid fuzz drivers automatically.
First, \FuDriver{} determines a set of entry functions and evaluates their respective chance of exhibiting a vulnerability.
Then, \FuDriver{} generates fuzz drivers for the entry functions through hierarchical parameter replacement and type inference. 
We implemented \FuDriver{} and evaluated its effectiveness on real-world programs selected from the Android Open-Source Project, Google's \textit{fuzzer-test-suite} and industrial collaborators. 
% \FuDriver{} can cover 79.54\% more branches, 105.52\% more paths, and detect 10 more unique bugs than the state-of-the-art fuzz driver synthesizer FUDGE and \% more branches, \% more paths over another state-of-the-art tool FuzzGen.
\FuDriver{} covered on average 1.08$\times$-2.03$\times$ more basic blocks and 1.36$\times$-2.06$\times$ more paths over state-of-the-art fuzz driver synthesizers \FUDGE{} and \FUZZGEN{}. 
\FuDriver{} performed on par with manually written drivers %and can perform better overall when also considering drivers for entry functions untouched by manual labor
and found 10 more bugs.
\end{abstract}


\section{Introduction}
\label{sec:introduction}
  
% Fuzz很NB AFL的简介
Fuzzing is a popular software testing technique for vulnerability detection. It attempts to trigger bugs within the program by generating massive amounts of input and monitoring the program's runtime state. Fuzzers have been able to find numerous vulnerabilities within real-world applications and are of great interest in both academia and industry.
AFL~\cite{afl} and LibFuzzer~\cite{libfuzzer} are two widely used fuzzers. 
Both use the classic genetic mutation algorithm to search for inputs that can improve coverage.
Due to AFL's popularity, there have been much research to improve its efficiency. Notable examples include AFLFast~\cite{bohme2017coverage}, FairFuzz~\cite{lemieux2018fairfuzz}, 
% 删QSym加MOpt
MOpt~\cite{lyu2019mopt},
%QSYM~\cite{yun2018qsym}, 
Angora~\cite{8418633} and Matryoshka~\cite{10.1145/3319535.3363225}.

%2. 驱动难以生成（现存的工作&现存工作的问题&实验结果）
Though fuzzing has achieved significant progress, there are still areas that require intensive manual labor, one of which is constructing effective fuzz driver programs for standalone libraries. This process usually requires the programmer to have a deep understanding of the program's source code, which is time-consuming and error-prone. Thus automating this process is imperative to improving the effectiveness of fuzzing.

% FUDGE/FuzzGen介绍 & 挑战
There have been some work into automated fuzz driver generation. Google recently proposed a framework \FUDGE{}~\cite{Wei2019FUDGE} to generate fuzz drivers semi-automatically. 
\FUDGE{} constructs fuzz drivers by scanning the program's source code to find vulnerable function calls and generates fuzz drivers using parameter replacement.
It works well in specific scenarios but tends to generate excessive fuzz drivers for large projects which require manual removal of invalid results.
Furthermore, it is infeasible to try all candidate drivers. 
Another example would be \FUZZGEN{}~\cite{251548}, which leverages a whole system analysis to infer the library's interface and synthesize fuzz drivers based on existing test cases accordingly. Its performance relies heavily upon the quality of existing test cases. 

In production environments, we face two significant challenges when constructing fuzz drivers automatically:
(1) One is to locate high-value entry functions. 
A high-value entry function should have the ability to reach lower levels of the program, yielding more code coverage after running for an extensive period. 
In addition, an entry function that contains vulnerable operations such as \textit{memcpy()} is considered to be more vulnerable and should require more attention.
(2) The other is to synthesize a valid and effective fuzz driver based on the identified entry function.
A correct fuzz driver should be able to call the target function with suitable parameters.

% 我们的设计：如何解决挑战 && 结果
In this paper, we present \FuDriver{} to address these challenges and construct fuzz drivers automatically. 
\FuDriver{} works as follows.
First, it scans the target program, looking for functions with high vulnerability priority as potential entry functions. Then, it synthesizes parameters using algorithmic methods. Finally, \FuDriver{} constructs the fuzz driver which calls the entry function with Address-Sanitizer~\cite{ASAN} enabled to initialize the fuzzing process.
% remove state-of-the-art
To evaluate its effectiveness, we test \FuDriver{} on some real-world projects selected from Android Open-Soure Project, Google's \textit{fuzzer-test-suite} and industrial collaborators. 
The results show that \FuDriver{} covers 1.08$\times$-2.03$\times$ more basic blocks, 1.36$\times$-2.06$\times$ more paths, and detects 10 more bugs than the  fuzz driver synthesizer \FUDGE{} and \FUZZGEN{} in total. Compared with the drivers written manually by domain experts, \FuDriver{} covers almost the same number of branches, paths, and bugs.  %for the same entry function. When accompanied by the synthesized fuzz drivers for other entry functions, \FuDriver{} outperforms the effectiveness of manually written drivers and can cover more branches, paths, and discover more bugs.

% 主要贡献
In summary, we make the following contributions: % in this papaer
\begin{itemize}
% 方法
\item{We propose a new fuzz driver synthesis framework, \FuDriver{}, which locates vulnerable functions and constructs fuzz drivers automatically, reducing the total amount of manual intervention required.} % add "and", "total"
% 技术实现
\item{We implement \FuDriver{} using the LLVM framework with generalized entry function location and accurate parameter inference, allowing for increased performance and compatibility than the state-of-the-art.
}  
% 对已有工作的改进
\item{We test \FuDriver{} on real-world projects selected from the Android Open-Soure Project, Google's \textit{fuzzer-test-suite} and projects from industrial collaborators. The result shows that \FuDriver{} covers 1.08$\times$-2.03$\times$ more blocks, 1.36$\times$-2.06$\times$ more paths and detects 10 more bugs than \FUDGE{} and \FUZZGEN{} and performs at least as well as manually written drivers.}
\end{itemize}

The rest of this paper is organized as follows. Section~\ref{sec:RelatedWork} introduces related work. 
Section~\ref{sec:Design} explains \FuDriver{}'s design, including the \EFL{} and the \FDS{}.
Section~\ref{sec:Implementation} covers the implementation details. 
Section~\ref{sec:Evaluation} evaluates \FuDriver{} on real-world programs and compares its results against \FUDGE{} and \FUZZGEN{}. % remove effectiveness
Section~\ref{sec:Studies} demonstrates the application in real industrial  practices. 
Section ~\ref{sec:LessonLearned} discusses the problems we encountered and potential future work. 
We conclude in Section \ref{sec:Conclusion}.

\section{Related Work}
\label{sec:RelatedWork}
  % (1) EnFuzz (2) SAFL (3) Polar
% (4) EMFuzz - 对固件进行模糊测试
% (5) ICS Protocal Fuzz : Peach* 是对Peach的改进
% 没引用Peach, Peach*和Polar等一起都堆在工业应用里
% (6) Fuzz Testing in practice 也是个工业应用 也堆在一起 

In this section, we introduce related work on fuzzing. We mainly discuss fuzzing in industry, automatic fuzz driver generation, API usage analysis and unit test generation.

% 总分总

\textbf {(1) Fuzzing in industry.} 
Fuzzing is a powerful technique for detecting vulnerabilities. 
% 加一句现有的工作?
% 很多工作都提升了fuzz的性能
There have been much research effort in improving fuzzing performance. InteFuzz~\cite{liang2019engineering}, DeepFuzz~\cite{liang2019deepfuzzer} and SAFL~\cite{wang2018safl} improves the overall efficiency of fuzzing algorithms. Zeror~\cite{zhou2020zeror} increases the fuzzing throughput by optimizing the fuzz target's execution speed. EnFuzz~\cite{chen2019enfuzz} and PAFL~\cite{liang2018pafl} allows fuzz engines to scale better.
% 上面是新加的
The effectiveness of the aforementioned projects have been demonstrated in numerous industrial applications \cite{shi2019industry, luo2019polar, gao2020fuzz, luo2020ics, liang2018fuzz, fu2019evmfuzzer}. 

Some other work have been conducted to streamline the fuzzing process.
For instance, Google’s OSS-Fuzz~\cite{ossfuzz}, which uses libFuzzer~\cite{libfuzzer} and AFL~\cite{afl} as its backend, has found thousands of bugs over a period of 5 months. 
ClusterFuzz~\cite{clusterfuzz} is the distributed infrastructure behind OSS-Fuzz, which automatically builds and executes binaries with different versions and finds the version that introduces a specific bug.
% AFL~\cite{afl} and LibFuzzer~\cite{libfuzzer} are two popular fuzzers widely deployed in industry. 
% ClusterFuzz uses AFL or LibFuzzer as its fuzzing engines. 
AFL targets the entire executable, generates random inputs for the program and monitors its runtime state. 
LibFuzzer fuzzes library functions by interfacing through the function \textit{LLVMFuzzerTestOneInput()}. 
It generates random data for this function and checks whether the program crashes during execution. 
Research based on LibFuzzer has led to a number of improved implementations, such as HonggFuzz~\cite{honggfuzz}. %, UniFuzzer~\cite{uniFuzzer} and BPF Fuzzer~\cite{bpfFuzzer}.

\textbf {(2) Automatic fuzz driver generation.}
Fuzz drivers are required when one wishes to fuzz a library function. 
Originally, testers wrote fuzz drivers for the target library by hand, which is inefficient and error-prone.
To synthesize fuzz drivers automatically, Google proposes a framework named \FUDGE{}~\cite{Wei2019FUDGE}. 
It scans the source code of the project for the vulnerable API, synthesizes the interface function \textit{LLVMFuzzerTestOneInput()} which calls the vulnerable API, and fuzzes it using LibFuzzer~\cite{libfuzzer}. \FUDGE{} operates by finding a function A, which calls another function B with the signature \textit{(uint8\_t*, size\_t)}. Then it considers function A as the entry function and binds the buffer generated by the fuzz engine with function B's signature.
This process will generate a large number of candidate drivers. \FUDGE{} shows them to testers directly, allowing them to modify those drivers manually to guarantee correctness. \FUZZGEN{} \cite{251548} scans the source code of the target project, finds the dependency of each API function, and generates interfaces to call the API functions based on existing test cases. Without qualified test cases, it will fail to generate correct drivers.   %It is similar to the traditional test case generation techniques such as Randoop \cite{pacheco2007randoop} and EvoSuite \cite{fraser2011evosuite}. 
%FuzzGen scans the source code of the target project, finds the dependency of each API function, and generates interfaces to call the API functions in order \cite{251548}. It is similar to the traditional test case generation techniques such as Randoop.

% 这个可以少一点
% 删一些
\textbf {(3) API Usage Mining.} Similar to API usage mining, fuzz driver synthesis needs to identify meaningful entry functions. 
% QoM~\cite{tansalarak2003qom} evaluates code quality to analyze code snippets.
% XSnippet~\cite{sahavechaphan2006xsnippet} allows tester to query code snippets that are relevant to the current code part.
MAPO~\cite{zhong2009mapo} searches and mines for the most frequent called APIs in the target project. 
GrouMiner~\cite{nguyen2009graph} mines usage patterns based on the graph of the project. 
APIExample~\cite{wang2011apiexample} extracts usage examples with similar functions.
% MUSE~\cite{moreno2015can} makes an improvement over APIMiner~\cite{wang2011apiexample} by combining static slicing with clone detection and heuristics ranking.

%\begin{comment}
\textbf{(4) Unit-test generation.}
Unit-test generation is another close research area to fuzz driver synthesis. Kampmann et al. ~\cite{kampmann2019carving} presents a method to automatically extract parameterized unit tests from system test executions. 
Testful ~\cite{baresi2010testful} generates test cases for Java classes and methods. 
Pacheco et al. ~\cite{pacheco2007feedback} presents a technique that improves random test generation by incorporating feedback obtained from executing test inputs as they are created.
GRT ~\cite{ma2015grt} uses static and dynamic analysis to include information on program types, data, and dependencies in various stages of automated test generation. 
% Rostra ~\cite{xie2004rostra} de-duplicates redundant test cases. 
GenRed ~\cite{jaygarl2010genred} generates and reduces object-oriented test cases. 
%\end{comment}

\begin{comment}
% 和之前两个项目的区别也说一下
\textbf{(5) Main differences.} Unlike other work, our research does not extract API functions directly or synthesize simple unit tests. Instead, we focus on locating vulnerable entry functions and synthesizing valid fuzz drivers for C/C++ projects.
FUDGE usually locates too many entry functions and generates too many small fuzz drivers for a given entry function, and it is not easy to select efficient drivers from those generated drivers. We intelligently locate and measure entry functions by its vulnerability priority and automatically synthesize valid fuzz drivers for each entry function without the need of manual modification or selection. 
\end{comment}

\section{\FuDriver{} Design}
\label{sec:Design}
  % design已经初步修改完毕

% 3.1 ：Ovewview放在Design里 Entry Function Locator 和 Driver Generator 这两个框突出 尽量都3个图。一到两段解释这个图

% 3.2 3.3介绍Locator和Generator 所有字黑体加粗 所有字加粗

% 所有图的解释减少 算法解释增多 

\begin{figure}[!htbp]
 \centering
 \includegraphics[width=0.49\textwidth]{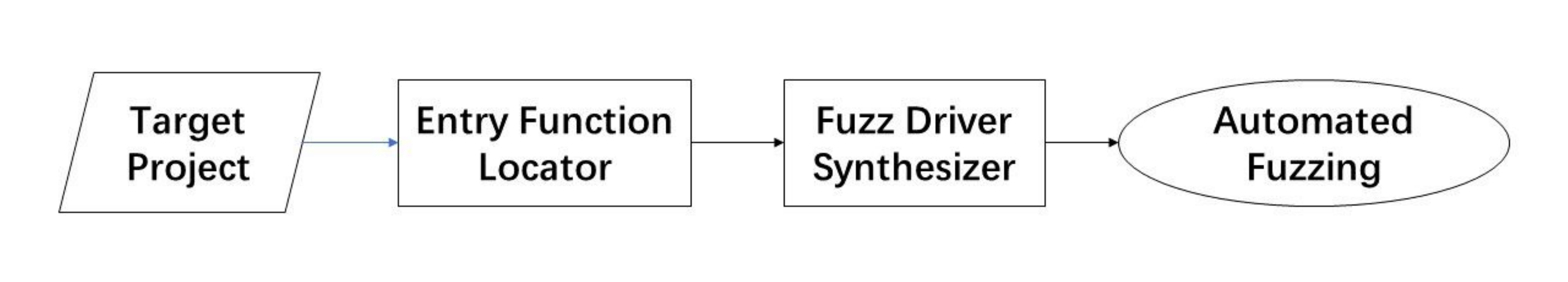}
 %\vspace{-0.5 cm}
 \caption{Overview of \FuDriver{}. It uses the \EFL{} to identify potential functions to fuzz and leverages the \FDS{} to construct fuzz drivers.}
 \label{fig:overview}
\end{figure}

Figure \ref{fig:overview} shows the high-level architecture of \FuDriver{}'s design. \FuDriver{} consists of two main modules: \EFL{} and \FDS{}. \EFL{} locates and sorts the vulnerable functions that contain many vulnerable operations such as \textit{memcpy()} in the target project. 
% FDS为EFL定位的入口函数合成参数并确保在程序运行过程中，由FuDriver生成的参数上不会发生buffer overflow
\FDS{} synthesizes arguments for the entry functions located by \EFL{} and ensures there will be no memory safety issues regarding the entry function arguments.
% \origin{The fuzz driver synthesizer mutates the arguments of the sub-functions, which the entry function calls, synthesizes parameters for the entry function, and builds function \textit{LLVMFuzzerTestOneInput(}\texttt{const uint8\_t*, size\_t}\textit{)}, which is the interface of LibFuzzer.} 

% Overview

\subsection{Entry Function Locator}

%入口程序很重要
The effectiveness of fuzz testing relies upon the quality of the entry function of the driver program.
An effective entry function may in turn call numerous other functions, potentially leading to higher edge coverage.
% \origin{An efficient entry function can cover most of the core codes in the project.}
In contrast, an ineffective driver may perform many checks on the supplied arguments, reducing the overall efficiency of fuzz testing
% \origin{Also, some entry functions may contain many error checks, which will reduce the overall efficiency of fuzz testing. }
A powerful fuzz driver should allow us to bypass those checks and reach the actual program logic directly.
%\origin{A useful entry function should bypass as much input checks as possible and be able to reach a significant proportion of the core code.}

% 为了Fuzz程序，程序员需要阅读源代码并寻找核心api
To Fuzz a project, testers need to communicate with its developers or read the relevant documents to manually find potential entry functions and construct valid function calls. 
This cannot guarantee quality drivers and is highly time consuming.
To reduce manual labor, \FUDGE{} and \FUZZGEN{} have been developed to synthesize fuzz drivers automatically. \FUDGE{} looks for functions with parameters \textit{(uint8\_t*, size\_t)} and assigns them with the buffer produced by fuzz engine. \FUZZGEN{} extracts API function dependency from the test cases and constructs a function to call API functions in a special order.
%目前没有人做入口定位
% \origin{Currently, there is a little research conducted regarding finding suitable entry functions in a target program.
% Before fuzzing a project, the testers need to communicate with the corresponding developers or read the corresponding documents in order to select potential entry functions manually. 
% Apparently, this method cannot guarantee high quality drivers while being highly time consuming.}
% FUDGE和FuzzGen的问题
However, \FUDGE{} may produce many ineffective drivers thus requiring manual intervention to pick and update effective fuzz drivers. \FUZZGEN{} on the other hand depends on the quality of the project's test cases. Without them, \FUZZGEN{} is incapable of generating any fuzz drivers.
% 如何解决问题
% \origin{To solve this problem, Google proposes a fuzz driver generation framework FUDGE.
% However, FUDGE has two significant drawbacks:
% \textit{(1) It may ignore some crucial functions.} 
% For example, functions with a local array can be vulnerable, but FUDGE does not specialize fuzz drivers for these functions. 
% \textit{(2) It may also locate excessive entry functions.} 
% FUDGE's pattern based selection criteria results in many ineffecient fuzz drivers generated.}

% FuDriver定位图
% \begin{figure}[!htbp]
%  \centering
%  \begin{tikzpicture}[node distance = 2.5cm]
% \node (input) [procedure] {Input Files};
% \node (extractast) [procedure, rounded corners, right of=input] {Extract AST};
% \node (sort) [procedure, rounded corners, right of=extractast] {Sort by \\Priority};
% \node (op) [procedure, above of=sort, yshift=-1cm] {Dangerous\\ Operations};
% \node (func) [procedure, below of=sort, yshift=1cm] {Entry\\ Functions};
% \draw [arrow] (input) -- (extractast);
% \draw [arrow] (extractast) |- (op);
% \draw [arrow] (op) -- (sort);
% \draw [arrow] (sort) -- (func);
% \end{tikzpicture}
%  %\vspace{-0.5 cm}
%  \caption{\FuDriver{}'s process for locating entry functions.}
%  \label{fig:locate}
% \end{figure}

\begin{figure}[htbp]
    \center
    \includegraphics[width=\linewidth]{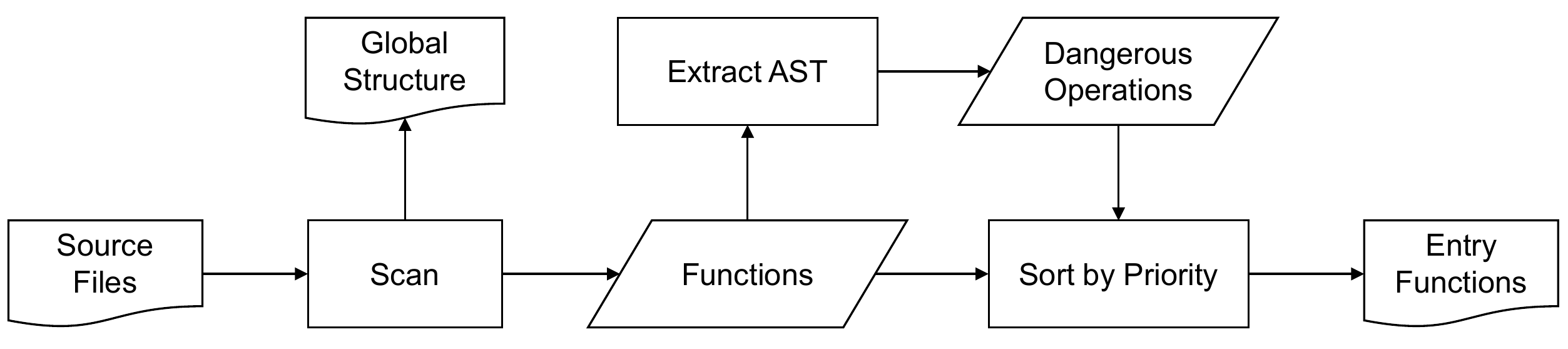}
    \caption{\FuDriver{}'s process for locating entry functions.}
    \label{fig:locate}
\end{figure}

% 解决问题
To solve these problems and automate the driver generation process, \FuDriver{} considers all functions as potential entry functions, evaluates their vulnerability priority and selects those with the highest priority as entry functions. Figure \ref{fig:locate} gives an overview of \FuDriver{}'s \EFL{}. 
% \origin{First, \FuDriver{} scans the abstract syntax tree (AST) of the project and extracts all functions found. To acquire the most suitable entry functions, \FuDriver{} evaluates the priority of each function based on the degree and frequency of vulnerable operations, then selects the functions with the highest priority as the entry functions.}

First, \FuDriver{} scans the project and extracts all functions found.
% and records all global structures. 
To acquire the most suitable entry functions, \FuDriver{} calculates the priority of each function based on the number of memory dereferencing operations.
% 新加的
% Algorithm \ref{alg-Priority} describes how we get an entry function's priority.

% 算法——优先度
\begin{algorithm}[!h]
\caption{Algorithm for evaluating potential entry functions' priorities.}
\label{alg-Priority}
\KwIn {source code files \textit{fileList}}
\KwOut {functions with their priority \textit{funcMap}}
\LinesNumbered
\textit{funcMap} = Map$<$Function, Priority$>$\\
\ForEach{\textit{file} \textnormal{in} \textit{fileList}} {
    \ForEach{\textit{func} \textnormal{in} \textit{file}} {
        \If {\textit{func} \textnormal{in} \textit{funcMap}} {
            continue\
        }
        \textit{funcMap}.push(\textit{func}, 0)\\
        \textit{priority} = 0\\
        \ForEach {\textit{stat} \textnormal{in} \textit{func}} {
            \uIf {\textit{stat} \textbf{dereferences} \textnormal{a pointer}} {
                \textit{priority} += 1\
            }
            \uElseIf {\textit{stat} \textbf{processes memory}} {
                \textit{priority} += 1\
            }
            \ElseIf {\textit{stat} \textnormal{calls} \textit{func2}} {
                \textit{priority} += \textit{func2}.getPriority()\
            }
        }
        \textit{funcMap}.update(\textit{func}, \textit{priority})\\
    }
}
\Return \textit{funcMap}\
\end{algorithm}

% 算法——危险度

%\origin{Algorithm \ref{alg2} presents our method of calculating the priority for each function. It takes the target function \textit{Function} as its parameter, and returns the function's priority.
%The algorithm first initializes the priority of the function to 0. Then, \FuDriver{} scans all parameters of the function, and increments the priority by one if a parameter is a \texttt{pointer}, as presented in Lines 3-7. 
%Next, \FuDriver{} performs depth first search on the AST \textit{FunctionAST} and counts the vulnerable operations that are prone to memory vulnerabilities. We regard the following types of operations as vulnerable operations:} % remove three
Algorithm \ref{alg-Priority} presents our method of accessing the priority for each function. It takes the list of all source files \textit{fileList} as its argument and returns \textit{funcMap}, a map between the functions and their priorities.
% 这地方的逻辑？
The algorithm operates by scanning all the input files. When it encounters a new function, it first initializes its priority to 0 and inserts it into \textit{funcMap}, as shown in line 2-7. The algorithm then scans all statements in the function to get its priority, as shown in line 8-17. We regard the following types of statements as potential vulnerable statements:

\begin{itemize}
    \item{\textbf{Dereferencing a pointer}. Buffer overflows are the most common security issues in industrial situations. These can only be triggered by dereferencing an invalid pointer.}
    \item{\textbf{Calling a memory related function}. Some \textit{libc} library functions such as \textit{memset()} and \textit{memcpy()} may dereference pointers internally. These functions are a potential source of buffer overflows.}
    \item{\textbf{Calling other functions in the same project.} If the child function can potentially cause buffer overflows, then the parent function should also be considered as vulnerable.}
\end{itemize}

\begin{comment}
\begin{itemize}
\item{
\textbf{Reading from or writing to memory directly.} Memory related operations could overflow, thus resulting in a memory vulnerability, which is most common in real-world software.
} 
\item{
\textbf{Calling memory related library functions.} Calling \textit{malloc()} or \textit{calloc()} without calling \textit{free()} afterwards may lead to memory leaks. Calling \textit{free()} inappropriately may lead to Double-Free or Use-After-Free vulnerabilities. Some functions (such as \textit{memcpy()}, \textit{memset()}, \textit{memmove()}, \textit{strcpy()}, etc.) read data from or write data to memory, which, when used incorrectly, may also cause buffer overflows.
} 
\item{
\textbf{Calling other functions in the target project.} If function \textbf{\textit{A}} calls function \textbf{\textit{B}}, the priority of function \textbf{\textit{A}} should be incremented with the priority of function \textbf{\textit{B}}.
}
\end{itemize}
\end{comment}

A function that contains a greater amount of vulnerable statements should receive a higher priority. After processing all statements in the function, its priority value in \textit{funcMap} is updated accordingly, as shown in line 18. Finally, the algorithm returns \textit{funcMap}, the priorities of all functions, as shown in line 21.

% 所有图的解释合成一段话

% 算法——定位

% \origin{The procedure of identifying entry functions is presented in Algorithm \ref{alg1}. The algorithm takes three parameters, namely \textit{EntryList} representing user-defined entry functions, \textit{MaxNumber} specifying the max number of entry functions, and \textit{FileList} containting the target program's source files.}
% the input files of the target program -> the target program's source files

Then, \FuDriver{} selects the functions with the highest priority as the entry functions.
The algorithm of the \EFL{} is presented in Algorithm \ref{alg-EntryFunctionLocator}. It takes two parameters, namely \textit{maxNumber}, specifying the max number of entry functions, \textit{fileList}, containing the target program's source files, and returns \textit{funcList} as entry functions. 
% 解释代码
First, \EFL{} evaluates all functions in the \textit{fileList} by their respective vulnerable priorities, as shown in lines 1-6. Then, it sorts the functions by their priorities, as shown in line 7. Next, it retrieves the \textit{maxNumber}-highest-priority functions as the final entry functions, as shown in lines 8-14.
% \origin{As presented in lines 4-8, the algorithm adds the user-defined entry functions, if any, to \textit{FuncCandidate} with the highest priority.
% Then, \FuDriver{} scans the source files specified by \textit{FileList} and builds ASTs accordingly, as shown in Line 11.
% Next, \FuDriver{} performs depth first search on each AST, evaluates the vulnerability priority for each function, and adds the function with its priority into the map \textit{FuncCandidate}, as shown from Line 12 to Line 20. 
% Considering that the \textit{main()} function and the test cases of the project often take fewer parameters and contain a complete procedure, \FuDriver{} assigns a higher priority for these functions.
% Finally, in Lines 22-27, the map \textit{FuncCandidate} is sorted according to the priority, and the entry functions with the highest priority are selected as the final entry functions. }

% 定位算法
\begin{algorithm}[!t]
\caption{Algorithm of \EFL{}} 
\label{alg-EntryFunctionLocator}
\KwIn {maximum number of entry functions \textit{maxNumber}\ \quad\quad\quad source code files \textit{fileList}}
\KwOut {entry functions \textit{funcList}}
\LinesNumbered
\textit{funcMap} = Map$<$Function, Priority$>$\\
\ForEach {\textit{file} \textnormal{in} \textit{fileList}} {
    \ForEach {\textit{func} \textnormal{in} \textit{file}} {
        \textit{funcMap.push(func.getPriority(func))}\
    }
}
\textit{funcMap} = \textit{funcMap}.sortBy(Priority)\\
\textit{funcList} = []\\
\While {\textit{maxNumber} $>$ 0 \textnormal{and not} \textit{funcMap}.empty()} {
    \textit{firstFunc} = \textit{funcMap}.pop()\\
    \textit{funcList}.push(\textit{firstFunc})\\
    \textit{maxNumber} -= 1\\
}
\Return \textit{FuncList}\
\end{algorithm}

Compared with \FUDGE{}'s pattern matching and \FUZZGEN{}'s API function searching in existing test cases, \FuDriver{}'s design is guarenteed to be general and versatile. \FuDriver{} can accurately identify vulnerable functions without resorting to additional information.

\subsection{Fuzz Driver Synthesizer}

% 驱动也很重要
The effectiveness of fuzz testing is highly dependent upon whether the driver is capable of invoking the entry function correctly.
% \origin{An optimal driver can pass the input generated by the fuzz engine to the target program correctly, allowing input mutations to effectively modify program behavior.}
An optimal driver can transform the input generated by the fuzz engine to the target function's arguments correctly.
Conversely, a faulty driver may even introduce errors, seriously affecting the performance of fuzzing.
Therefore, synthesizing high-quality fuzz drivers based on the located entry functions is vital towards good fuzzing performance. 

% 介绍LibFuzzer
Fuzz drivers are tightly coupled with the fuzzing engines. \LIBFUZZER{} is a widely-used fuzzing engine and has been applied to find many serious bugs in previous evaluations. Therefore, we choose \LIBFUZZER{} as \FuDriver{}'s fuzzing engine. \LIBFUZZER{} uses the interface function \textit{LLVMFuzzerTestOneInput()} to initiate its fuzzing process, so \FuDriver{} synthesizes this interface function for the located entry function.

% 图要简单 算法要详细

\begin{figure}[!htbp]
 \centering
 \includegraphics[width=0.6\linewidth]{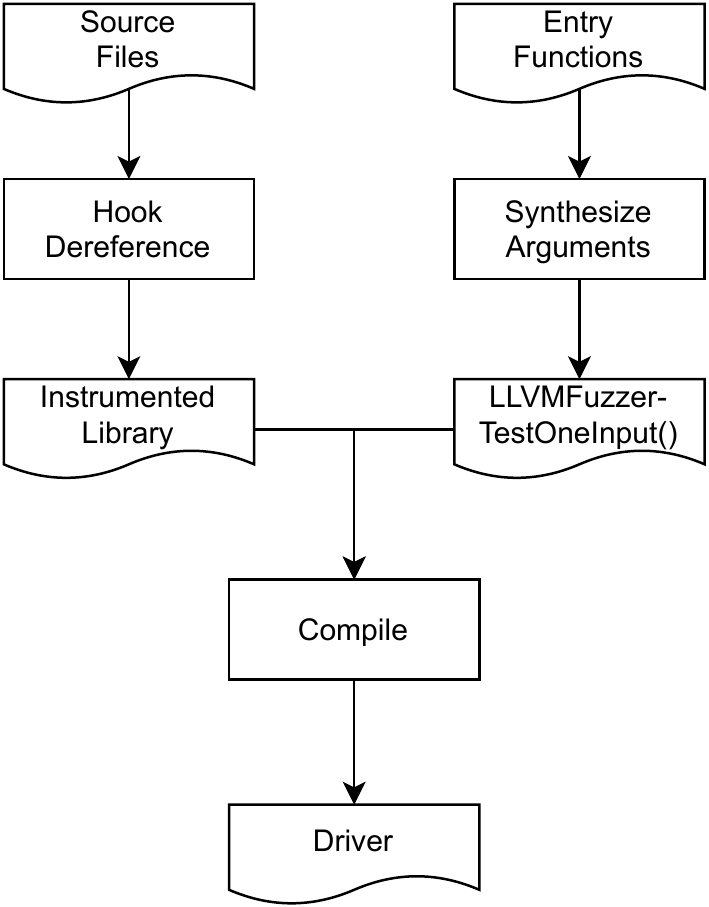}
 %\vspace{-0.5 cm}
 \caption{\FDS{}'s process of fuzz driver synthesis.}
 \label{fig:synthesize}
\end{figure}

\begin{algorithm}[!t]
\caption{Runtime Value Assignment Routines. Functions \textit{getScalarValue()} and \textit{getPointerValue()} are inserted to assign values for scalar and pointer objects. Functions \textit{LoadInstHook()} and \textit{StoreInstHook()} are inserted before each load and store instruction.}
\label{alg-FuzzDriverSynthesizer}
% \KwIn {The dereferenced pointer value $ptr$}
% \KwOut {The modified pointer value $ptr\_ret$}
\SetKwFunction{FMain}{...}
\SetKwProg{Fn}{Function}{:}{}
\LinesNumbered
% \textit{driver} = emptyLLVMFuzzerTestOneInput()\\
% \textit{arguments} = []\\
\begin{comment} 
\ForEach{\textnormal{argument} \textit{arg} \textnormal{in} \textit{func}} {
    \uIf {\textit{arg} \textnormal{is} \textbf{scalar number}} {
        \textit{arg} = getScalarValue()\\
    } \uElseIf {\textit{arg} \textnormal{is} \textbf{pointer}} {
        \textit{arg} = getPointerValue()\\
    } \ElseIf {\textit{arg} \textnormal{is} \textbf{struct} \textnormal{or} \textbf{array}} {
        \ForEach{\textit{subarg} \textnormal{in} \textit{arg}} {
            getArg(\textit{subarg})
        }
        \textit{arguments}.push(\textit{arg})
    }
}
\end{comment}
\Fn{\textnormal{getScalarValue}(size)}{
    \textbf{return} readFromBuffer(\textit{size})\;
}
\textbf{endFunction}\\
\Fn{\textnormal{getPointerValue}()}{
    \textit{ptr} = malloc(SIZE)\;
    markAsUnassigned(\textit{ptr}, SIZE)\;
    \textbf{return} \textit{ptr}\;
}
\textbf{endFunction}\\
\Fn{\textnormal{LoadInstHook}(ptr)} {
    \If {\textnormal{NotAssigned}(ptr)} {
        \tcc{\textit{Compile-time determined}}
        \uIf {\textnormal{isScalarType}} {
            *\textit{ptr} = getScalarValue(OBJECT\_SIZE)\;
        } \uElseIf {\textnormal{isPointerType}} {
            *\textit{ptr} = getPointerValue()\;
        } \ElseIf {\textnormal{isStructOrArrayType}} {
            \textit{recursively assign all members} \;
        }
        markAsAssigned(\textit{ptr})\;
    }
}
\textbf{endFunction}\\
\Fn{\textnormal{StoreInstHook}(ptr)} {
    markAsAssigned(\textit{ptr})\;
}
\textbf{endFunction}\\
\begin{comment}
\Fn{getStructOrArrayValue(ptr)}{
    \ForEach{\textit{subPtr} \textnormal{in} \textit{ptr}} {
        initPtr(subPtr);
        \uIf {typeof(\textit{subPtr}) \textnormal{is} scalar} {
            
        } \uElseIf {typeof(\textit{subValue}) \textnormal{is} pointer} {
        
        } \ElseIf {typeof(\textit{subValue}) \textnormal{is} struct}
    }
}
\end{comment}
% \Fn{initValue(value)}

%\ForEach {\textit{stat} \textnormal{in} \textit{func}} {
%     \uIf {\textit{stat} \textbf{stores} \textnormal{into marked memory} \textit{mem}} {
%         markAsStore(\textit{mem})\
%     } \uElseIf{\textit{stat} \textbf{loads} \textnormal{from marked memory} \textit{mem}} {
%         \If {\textit{mem} \textnormal{has NOT been stored}} {
%             \textit{type} = typeof(\textit{stat})\\
%             \textit{param} = getParam(\textit{type})\\
%             \textbf{store} \textit{param} into \textit{mem}\\
%             markAsStore(\textit{mem})
%         }
%     } \ElseIf {\textit{stat} \textbf{processes memory}} {
%         expand \textit{stat} to \textbf{store} and \textbf{load} statements\\
%         do as mentioned above\\
%     }
%}

\end{algorithm}

Figure \ref{fig:synthesize} shows how \FuDriver{} synthesizes a fuzz driver in detail. It takes the entry function as its input, and returns a fuzz driver in the form of the \textit{libFuzzer} interface function \textit{LLVMFuzzerTestOneInput()}.
First, \FuDriver{} builds an empty function \textit{LLVMFuzzerTestOneInput()}. Then, \FuDriver{} instruments the entry function's arguments to assign valid values at runtime. Specifically, it analyzes the argument's value type % \FuDriver{} then generates arguments 
using the following rules:
\begin{itemize}
    \item {\textbf{Scalar Types}: For scalar types, such as integers and floating point numbers, \FuDriver{} instruments runtime instructions that read subsequent bytes from the input buffer generated by \LIBFUZZER{} and assign them to the argument value.} 
    % \FuDriver{} reads some segments from the buffer generated by \LIBFUZZER{} and assigns it to the argument.}
    % \item {\textbf{Floating point numbers}: \FuDriver{} maintains a random number generator, reads an integer from the fuzzing engine and generates a floating-point number based on integer.}
    \item {\textbf{Pointer Types}: For pointer types, \FuDriver{} instruments a runtime function that allocates memory without initialization and assigns the pointer value with its starting address. The size of the allocated memory should be large enough to accommodate either the object that the pointer points to or a small integer array.}
    % \FuDriver{} allocates a chunk of memory \textit{without filling anything into it} and assigns the memory to the argument. The chunk of memory should be large enough to accommodate either the object that the pointer points to or a small integer array.}
    \item {\textbf{Array or Structure Types}: For array or structure types, \FuDriver{} assigns values for its member values recursively by repeating this process on each member.}
\end{itemize}
Next, \FuDriver{} scans the entry function and inserts statements to \textit{lazy-stores} values into arguments with pointer type. 
Generally, we can not know the actual type of a pointer until it is dereferenced, because C allows casting any pointer to \textit{void*} and vice versa. 
Hence, we can not assign a fitting value to the pointer based on its current type. 
To solve this, we must know the type in which it is used, and one method to acquire this information is to trace dereference operations of the pointer. 
Therefore, we delay the assignment of the values referenced by pointers until its first usage, hence the name \textit{lazy-store}.
\FuDriver{} keeps track of whether a memory range has been previously assigned a value. 
All pointer values in the entry function's arguments are initially marked as \textit{unassigned}. 
For store operations, \FuDriver{} inserts instructions that mark the memory range as \textit{assigned}.
For load instructions, \FuDriver{} instruments a runtime function that checks if the memory range is marked as \textit{assigned}. 
If not, it assigns a value corresponding to the dereferenced type using the method for entry function arguments. 
When we find a \textit{load} operation and the memory that it points to has not been previously assigned, \FuDriver{} constructs a parameter with the same type using the method for constructing argument values and stores it into the relevant memory area just before loading from it. 
% Furthermore, to avoid assigning repeatedly to one chunk of memory, \FuDriver{} marks the memory as occupied when storing into it as shown in line 23 and 29 of Algorithm \ref{alg-FuzzDriverSynthesizer} and does not store into memory if the memory area is marked as occupied, as shown in line 25 of Algorithm \ref{alg-FuzzDriverSynthesizer}. 
We also expand memory related functions such as \textit{memcpy()} and \textit{memset()} into load and store operations for a more complete analysis. 
Algorithm \ref{alg-FuzzDriverSynthesizer} shows the instrumented functions that assign values in the situations described above during runtime.

\section{Implementation}
\label{sec:Implementation}
% TODO

% We implement FuDriver on LLVM!!!! 在这里写 然后大概写一下FuDriver工作流程
% Locator的实现 通过clang转成IR就完了
% 比如memory leak
% 比如如何排除无效驱动
% 大概想3-4个

% \origin{In this section, we introduce the implementation details of \FuDriver{}.
% Considering that almost all projects can be compiled into LLVM-IR, we can insert the interface function \textit{LLVMFuzzerTestOneInput()} generated by \FuDriver{} directly into the target's LLVM-IR to facilitate fuzzing. 
% We mainly solve the following challenges during implementation:
% Considering that almost all projects can be compiled into LLVM-IR, we can insert the interface function \textit{LLVMFuzzerTestOneInput()} generated by \FuDriver{} directly into the target's LLVM-IR to facilitate fuzzing. 
% We mainly solve the following challenges during implementation:}
In this section, we introduce the implementation details of \FuDriver{}. We implement \FuDriver{} using the LLVM compiler framework. The interface function \textit{LLVMFuzzerTestOneInput()} generated by \FuDriver{} is constructed at the Intermediate Representation (IR) level. \FuDriver{} links this function with the library code together into one bitcode file. We can then compile the bitcode file to an executable fuzz driver. We mainly solve the following challenges during implementation:

% priority
% \origin{\textbf{Evaluating the priority of functions.} When locating entry functions, we need to calculate the priority for each function. 
% In essence, we should count the number of memory operations and memory-related function calls. Because LLVM-IR uses the \textit{load} and \textit{store} instructions to read from and write to memory respectively, and \textit{call} or \textit{invoke} instructions to call a function, we can count the total number of \textit{load}, \textit{store}, \textit{call} and \textit{invoke} instructions to evaluate the priority of a function.}
\textbf{Evaluating the priority of functions.} When locating entry functions, we need to calculate their respective priorities. 
In essence, we should count the number of statements which dereferences a pointer or calls other functions to processes memory. LLVM-IR uses \textit{load} and \textit{store} instructions to read from and write to memory through a pointer respectively and \textit{call} instructions to call a function, so we can trace these three kinds of instructions to get the priority for any function.

% 选择哪些作为入口
\textbf{Choosing effective entry functions.} \FuDriver{}'s \EFL{} can locate a few potentially effective entry functions, but the user should have to choice to select which function to fuzz. \FuDriver{} shows the recommended entry functions and testers can manually intervene to determine which entry functions to use or generate drivers for all potential entry functions automatically.

% lazy-store
\textbf{Avoiding redundant memory assignments.} \FuDriver{} will only lazy-store into any memory area if the area has not been previously marked as occupied. To keep track of all occupied memory segments, \FuDriver{} maintains a global map and inserts an instruction to mark the corresponding memory as filled before a corresponding store instruction.

\textbf{Synthesizing complex arguments for the entry function.} An entry function may contain pointer-typed parameters, each pointing to a structure containing another complex structure. By using the lazy-store technique, \FuDriver{} can generate these arguments correctly. For any pointer arguments, as mentioned in Algorithm \ref{alg-FuzzDriverSynthesizer}, \FuDriver{} assigns a plain chunk of memory. \FuDriver{} will only store an object if it is ever dereferenced. For all member variables, they are generated recursively. This allows \FuDriver{} to handle parameter construction with ease.
% and another chunk of memory will be assigned to the sub-element of this variable. So \FuDriver{} can handle almost all kinds of parameters.

% 参数的参考值
\textbf{Assigning appropriate values for arguments.} Generating random values for an argument may decrease the effectiveness of fuzz testing. To find an appropriate value for an argument, \FuDriver{} scans the IR of the function in search for comparison instructions. If one of the operands used by a comparison instruction is an argument pending assignment, then the other operand will be considered as an appropriate value for the argument. \FuDriver{} will generate additional code to decide if the appropriate value should be assigned to the argument.

% 删掉多余的driver
% \origin{\textbf{Filtering out useless drivers.} 
% We can not guarantee that all synthesized drivers are valid, in particular the correctness of the synthesized parameters. 
% To evaluate a synthesized driver's effectiveness, we execute the driver for about 3 seconds automatically and monitor its runtime state. If it crashes, we consider the driver to be invalid. An invalid driver is often caused by API misuse where some parameters should be assigned with special values. If \FuDriver{} chooses a low-level function as its entry function and assigns a wrong value to its parameter, the driver will crash promptly. }
\textbf{Filtering out useless drivers.} 
Though \FuDriver{} uses many techniques to generate arguments for the entry function, We cannot guarantee that all synthesized drivers will be valid, especially for entry functions containing function pointer arguments. 
To evaluate a synthesized driver's effectiveness, we execute the driver for a short amount of time automatically and monitor its runtime state. If it crashes, we consider the driver to be invalid and it will be removed. 
An invalid driver is often caused by dereferencing a function pointer or an assertion statement failing. 
%Currently, \FuDriver{} cannot assign values for function-pointers reliably, so an assertion failure will force the driver exiting.
%
% \origin{Apart from API misuse, memory leaks may also invalidate a fuzz driver. 
% If an entry function allocates memory on the heap and returns its pointer, then a memory leak error will occur when running \FuDriver{} synthesized drivers. 
% To overcome this problem, \FuDriver{} first scans the program to find specific functions which take one pointer to a specific type as a parameter and call function \textit{free()} inside it. If \FuDriver{} can find such functions, it will also call the free function after calling the entry function; otherwise, \FuDriver{} will directly call \textit{free()} on the returned pointer.}
%\textbf{Keeping track of memory leaks.} 
Memory leaks may also invalidate a fuzz driver. If an entry function allocates memory on the heap but does not free it, then a memory leak error will occur. To avoid memory leaks, \FuDriver{} hooks memory allocation/deallocation functions such as \textit{malloc()} and \textit{free()} with \textit{\FuDriver{}\_alloc()} and \textit{\FuDriver{}\_free()}. \textit{\FuDriver{}\_alloc()} records the memory it allocates, and \textit{\FuDriver{}\_free()} remove the record of the memory it frees. At the end of function \textit{LLVMFuzzerTestOneInput()}, all memory recorded by \textit{\FuDriver{}\_alloc()} but not freed by \textit{\FuDriver{}\_free()} will be freed altogether.

\section{Evaluation}
\label{sec:Evaluation}
% TODO
% 第7页表排版可能要改

% 把最终的3个Driver放到一起，放到Case Study的开头，用一段话总的介绍一下，
% 然后再现在1 2 3分，细节【看着删除】。
% 增长不要从8000开始（前1分钟多采几个样）

% FuDriver-Total Fudriver-total
% 时间和bugs数目 换位置
% 全文In order to 全都改成to
% 所有could都改成can或者去掉

% 介绍这9个项目
% -------------------和FuzzGen对比 就不标颜色了
To examine the effectiveness of \FuDriver{}, we evaluate it on the 6 real-world libraries of Android Open-Soure Project (AOSP) used in the evaluation of \FUZZGEN{}, 9 real-world projects of Google's \textit{fuzzer-test-suite} with manually written drivers for the comparison of Google's FUDGE, and 3 real-world projects from industrial collaborators. 
%These libraries and projects are libraries with complex encode/decode functionalities, including \textit{libavc}, \textit{libhevc}, \textit{libgsm}, \textit{libmpeg2}, \textit{libopus} and \textit{libvpx}. 
These projects consist of image processing libraries (\textit{libjpeg}), file processing libraries (\textit{libxml2}, \textit{JSON}), regular expression engines (\textit{pcre2}), asynchronous resolver libraries (\textit{c\_ares}), font compression and decompression libraries (\textit{woff2}, \textit{libhevc}, \textit{libhavc}), and font shaping libraries (\textit{harfbuzz}).

We use two commonly used metrics to evaluate the effectiveness of automatic generated fuzz drivers on these real-world libraries, namely basic block coverage and path coverage. An effective driver should be capable of allowing the fuzz engine to cover a significant amount of the code. We collect the code coverage information using LLVM-cov, which provides us information on block coverage, and path coverage.

We conduct our experiments on a machine with Intel Xeon Gold 6148 processors and 128GiB of memory running on 64-bit Ubuntu Linux 18.04.
We run each fuzz driver 10 times, each using four threads over a period of 6 hours and report the average coverage statistics.

\subsection{Comparison with \FUZZGEN{}} 
We compare \FuDriver{} and \FUZZGEN{} on the effectiveness of synthesized drivers based on the six projects used in the evaluation of \FUZZGEN{}. 
For \FUZZGEN{}, we use the same fuzz driver provided with their project\footnote{\textit{libopus} and \textit{libvpx}'s sample drivers provided by \FUZZGEN{} are invalid and we could not execute them successfully.}. 
%If there are multiple drivers provided for one project, we then choose the most sophisticated valid fuzz driver.\footnote{Some examples provided by \FUZZGEN{} are invalid and we could not execute them successfully.}
For \FuDriver{}, we use \FuDriver{}'s \EFL{} to identify an initial set of entry functions, %We then select a final group of entry functions manually to reduce the amount of fuzz drivers generated. 
and utilize \FDS{} to synthesize a fuzz driver that calls the identified entry functions.
The results are presented in Table \ref{table-Coverage}, which shows the number of blocks, and paths. %functions, and lines of each driver covers respectively. %For page limit, we do not include path coverage, which is consistent to the branch coverage and can be referred to the website of \FuDriver{}.

\newcolumntype{C}{>{\arraybackslash}p{1.0cm}}
\NewEnviron{tableEnv}[1]{
  \begin{table}[!htbp]
    \caption{#1}
    \scriptsize
    \center
    \scalebox{1}[1]{%
      \label{table-Coverage}
      \begin{tabular}{p{0.9cm}p{1.3cm}|CC}
        \toprule
        {\mysize Project}
        & {\mysize }
        & {\mysize Blocks}
        & {\mysize Paths}\\
        \midrule
        \BODY
        \bottomrule
      \end{tabular}
    }
  \end{table}%
}{}

\begin{tableEnv}{Number of blocks and paths covered by fuzz drivers synthesized by \FuDriver{} and \FUZZGEN{}{}. We do not list the bugs found in these libraries since these do not have a standard bug list.}
\multirow{2}{*}{libavc}   & \FuDriver{} & 999  & 765 \\
                          & \FUZZGEN{}  & 579  & 81 \\
%                          & increase    & 0.725  \\
\midrule
\multirow{2}{*}{libgsm}   & \FuDriver{} & 1258 & 659 \\
                          & \FUZZGEN{}  & 1339 & 1173 \\
%                          & increase    & 0.006   \\
\midrule
\multirow{2}{*}{libhevc}  & \FuDriver{} & 928  & 476\\
                          & \FUZZGEN{}  & 558  & 188\\
%                          & increase    & 0.663  \\
\midrule
\multirow{2}{*}{libmpeg2} & \FuDriver{} & 1105 & 630\\
                          & \FUZZGEN{}  & 44   & 70\\
%                          & increase    & 24.1  \\
\midrule
\multirow{2}{*}{libopus}  & \FuDriver{} & 124  & 112\\
                          & \FUZZGEN{}  & -    & -  \\
\midrule
\multirow{2}{*}{libvpx}   & \FuDriver{} & 3221 & 924\\
                          & \FUZZGEN{}  & - & -      \\
\midrule
\multirow{3}{*}{total}    & \FuDriver{} & 7635 & 3566\\
                          & \FUZZGEN{}  & 2520 & 1512\\
                          & increase    & 2.03X$\uparrow$ & 1.36X$\uparrow$ \\
\end{tableEnv}

% 所有项目的图
\begin{figure*}[!htb]
    \centering
    %\subfigtopskip=2pt %设置子图与上面正文或别的内容的距离
    %\subfigbottomskip=2pt %设置第二行子图与第一行子图的距离，即下面的头与上面的脚的距离
    %\subfigcapskip=-5pt %设置子图与子标题之间的距离
    %\subfigure[basic block coverage] {
        %\includegraphics[width=0.5\textwidth]{NewImage/libavc-BasicBlock.pdf}
    %}
    %\subfigure[function coverage] {
        %\includegraphics[width=0.22\textwidth]{NewImage/libavc-Function.pdf}
    %}
    %\subfigure[line coverage]{
        %\includegraphics[width=0.22\textwidth]{NewImage/libavc-Line.pdf}
    %}
    %\vspace{-0.5 cm}
    \includegraphics[width=1.0\textwidth]{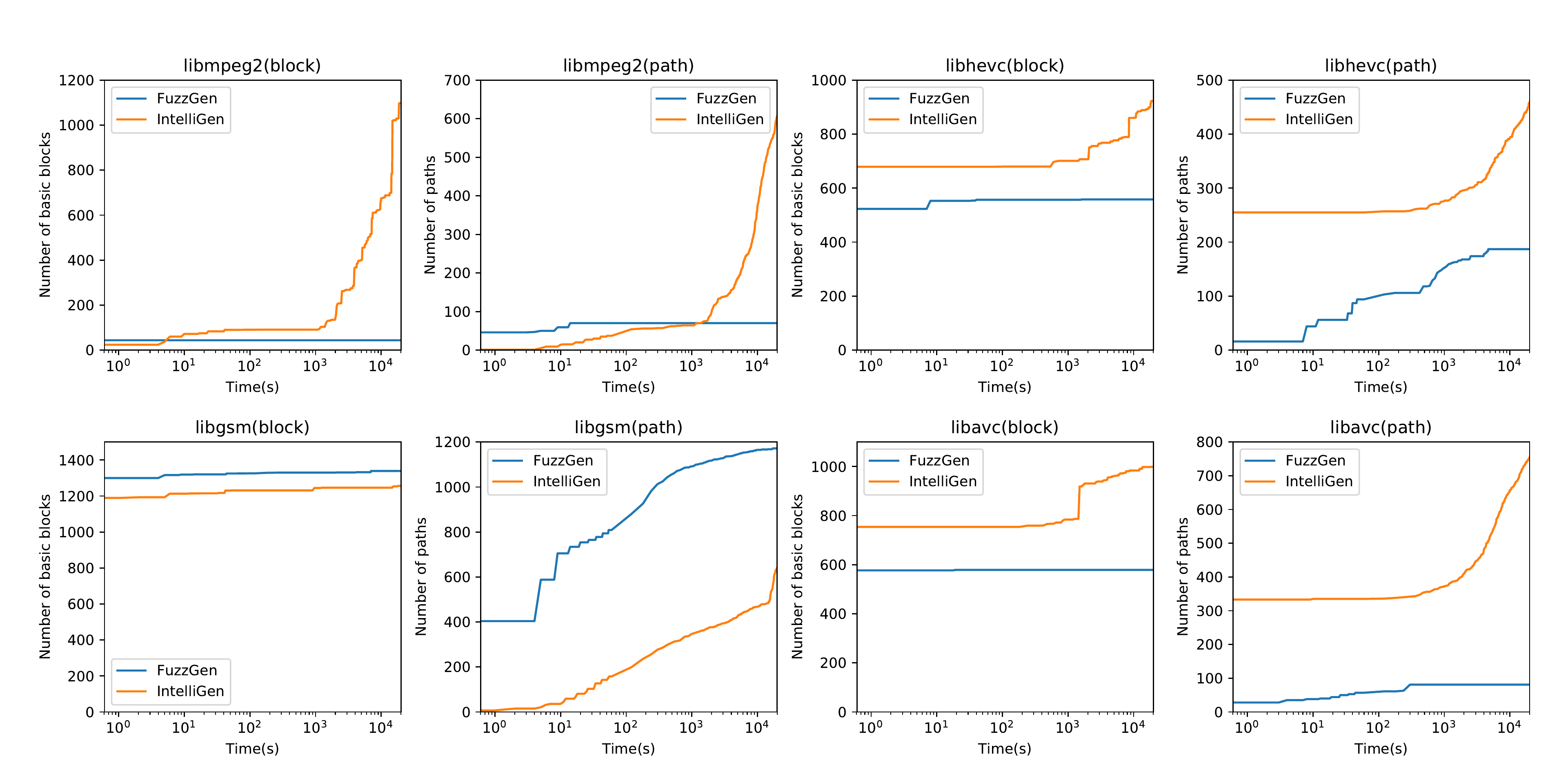}
    \label{comp-FuzzGen}
    \caption{Comparisons of the number of blocks and paths covered in \textit{libavc}, \textit{libgsm}, \textit{libhevc} and \textit{libmpeg2} by \FuDriver{} and \FUZZGEN{}. The drivers synthesized by \FUZZGEN{} for \textit{libopus} and \textit{libvpx} are invalid and not demonstrated in this figure. }
    % \label{fig-all}
\end{figure*}

As shown in Table \ref{table-Coverage}, \FuDriver{} is able to achieve better code coverage than \FUZZGEN{} on most projects, exceeding 50\% more block and path coverage on \textit{libavc} and \textit{libhevc} over \FUZZGEN{}. %Surprisingly, \FuDriver{} can cover 2000\% more basic blocks than \FUZZGEN{} on \textit{libmpeg2}. 
The only exception is \textit{libgsm}, where \FuDriver{} covers slightly less blocks than \FUZZGEN{} though covering more paths. The fuzz drivers provided by \FUZZGEN{} for \textit{libopus} and \textit{libvpx} are invalid, whereas \FuDriver{} can synthesize correct drivers for these two libraries. 
%
%Figure~\ref{comp-FuzzGen} 
Figure 4 demonstrates the effectiveness of the fuzz drivers synthesized by \FuDriver{} and \FUZZGEN{} over time. 
% To show the difference between \FuDriver{} and \FUZZGEN{} clearly, we plot time-coverage figure for each project. First of all, let us see Figure \ref{fig-libavc} and Figure \ref{fig-libhevc}, which shows the tendency of library \textit{libavc} and \textit{libhevc} respectively.
% 走势
% \begin{figure}[!htb]
%     \centering
%     \subfigtopskip=2pt %设置子图与上面正文或别的内容的距离
%     \subfigbottomskip=2pt %设置第二行子图与第一行子图的距离，即下面的头与上面的脚的距离
%     \subfigcapskip=-5pt %设置子图与子标题之间的距离
%     \subfigure[basic block coverage] {
%         \includegraphics[width=0.5\textwidth]{NewImage/libavc-BasicBlock.pdf}
%     }
%     \subfigure[function coverage] {
%         \includegraphics[width=0.22\textwidth]{NewImage/libavc-Function.pdf}
%     }
%     \subfigure[line coverage]{
%         \includegraphics[width=0.22\textwidth]{NewImage/libavc-Line.pdf}
%     }
%     %\vspace{-0.5 cm}
%     \caption{Code coverage for \textit{libavc}}
%     \label{fig-libavc}
% \end{figure}

% \begin{figure}[!htb]
%     \centering
%     \subfigtopskip=2pt %设置子图与上面正文或别的内容的距离
%     \subfigbottomskip=2pt %设置第二行子图与第一行子图的距离，即下面的头与上面的脚的距离
%     \subfigcapskip=-5pt %设置子图与子标题之间的距离
%     \subfigure[basic block coverage] {
%         \includegraphics[width=0.5\textwidth]{NewImage/libhevc-BasicBlock.pdf}
%     }
%     \subfigure[function coverage] {
%         \includegraphics[width=0.22\textwidth]{NewImage/libhevc-Function.pdf}
%     }
%     \subfigure[line coverage]{
%         \includegraphics[width=0.22\textwidth]{NewImage/libhevc-Line.pdf}
%     }
%     %\vspace{-0.5 cm}
%     \caption{Code coverage for \textit{libhevc}}
%     \label{fig-libhevc}
% \end{figure}

We examined the source code of the libraries we tested and analyzed the reason for \FuDriver{}'s better performance. In the \textit{libavc} and \textit{libhevc} libraries, \FuDriver{} and \FUZZGEN{} both synthesized fuzz drivers for the high level APIs (\textit{ih264d\_api\_function()} and \textit{ihevcd\_cxa\_api\_function()}. However, as \FuDriver{} is capable of constructing argument values using principled methods while \FUZZGEN{} is only capable of inferring possible values from test cases, the driver \FuDriver{} is more versatile and potentially capable of reaching more code.
A more prominent example is \textit{libmpeg2}. \FuDriver{} and \FUZZGEN{} are both capable of synthesizing fuzz drivers for the API function \textit{impeg2d\_api\_function()}. However, as the project does not provide test cases with much insight, \FUZZGEN{} cannot construct arguments that allow the API function to reach large proportions of the code, thus limiting the fuzz driver's performance. \FuDriver{}'s approach on the other hand allows for vastly better performance.

% \begin{figure}[!htb]
%     \centering
%     \subfigtopskip=2pt %设置子图与上面正文或别的内容的距离
%     \subfigbottomskip=2pt %设置第二行子图与第一行子图的距离，即下面的头与上面的脚的距离
%     \subfigcapskip=-5pt %设置子图与子标题之间的距离
%     \subfigure[basic block coverage] {
%         \includegraphics[width=0.5\textwidth]{NewImage/libmpeg2-BasicBlock.pdf}
%     }
%     \subfigure[function coverage] {
%         \includegraphics[width=0.22\textwidth]{NewImage/libmpeg2-Function.pdf}
%     }
%     \subfigure[line coverage]{
%         \includegraphics[width=0.22\textwidth]{NewImage/libmpeg2-Line.pdf}
%     }
%     %\vspace{-0.5 cm}
%     \caption{Code coverage for \textit{libmpeg2}}
%     \label{fig-libmpeg2}
% \end{figure}

Both \FuDriver{} and \FUZZGEN{} synthesized fuzz drivers on a single exposed API function for the previous libraries. In \textit{libgsm} however, there are multiple potential entry functions. \FuDriver{}'s \EFL{} identifies four entry functions and its \FDS{} constructs one driver each and one driver which calls all four functions consecutively, thus five drivers in total. \FUZZGEN{} also provides five drivers. Surprisingly, the best performing drivers for both \FuDriver{} and \FUZZGEN{} call the same function only.

Overall, \FuDriver{} performs better on the majority of libraries than \FUZZGEN{}, even though \FUZZGEN{} leverages additional information extracted from test cases while \FuDriver{} relies only on algorithmic parameter synthesis. 

\subsection{Comparison with FUDGE}

We also compare the performance of \FuDriver{} and \FUDGE{} regarding the effectiveness of synthesized drivers. %
We select the driver from \FUDGE{} with the highest performance out of the 100 sampled valid drivers. 
The results are presented in Table \ref{table-Branch-FUDGE} 
%, \ref{table-Path-FUDGE} and \ref{table-Bug-FUDGE}
, which show the number of blocks, the number of paths, 
and unique crashes of each driver respectively. 
%For page limit, we do not include path coverage, which is consistent to the branch coverage and can be referred to the website of \FuDriver{}.

% \newcolumntype{C}{>{\arraybackslash}p{1.0cm}}
\NewEnviron{mytable3}[1]{
  \begin{table}[!htbp]
    \caption{#1}
    \scriptsize
    \center
    \scalebox{1}[1]{%
      \label{table-Branch-FUDGE}
      \begin{tabular}{p{0.9cm}p{1.0cm}|CCCC}
        \toprule
        {\mysize Project}
        & {\mysize }
        & {\mysize Blocks}
        & {\mysize Paths}
        & {\mysize Bugs}\\
        \midrule
        \BODY
        \bottomrule
      \end{tabular}
    }
  \end{table}%
}{}

\begin{mytable3}{Number of blocks, paths covered and bugs found by \FuDriver{} and \FUDGE{} on Google's \textit{fuzzer-test-suite}.}
\multirow{2}{*}{re2}     & \FuDriver{}   & 2050  & 11228  & 0     \\
        & \FUDGE{} & 1203   &  100  & 0    \\
        \midrule
\multirow{2}{*}{harfbuzz}& \FuDriver{}   & 6259  & 12708  & 1   \\
        & \FUDGE{} & 82     & 3952    & 0      \\
        \midrule
\multirow{2}{*}{guetzli} & \FuDriver{}   & 692   &    7728    & 3    \\
        & \FUDGE{} & 80    & 1460    & 0        \\
        \midrule
\multirow{2}{*}{libjpeg} & \FuDriver{}   & 1236  & 780  & 0   \\
        & \FUDGE{} & 1059  & 738   & 0   \\
        \midrule
\multirow{2}{*}{woff2}   & \FuDriver{}   & 100    & 3936    & 2     \\
        & \FUDGE{} & 69    & 59    & 0       \\
        \midrule
\multirow{2}{*}{json}    & \FuDriver{}   & 903   & 3278   & 1    \\
        & \FUDGE{} & 546   & 632    & 1      \\
        \midrule
\multirow{2}{*}{libxml}  & \FuDriver{}   & 2355  & 128  & 1     \\
        & \FUDGE{} &  1529  &  12  & 0     \\
        \midrule
\multirow{2}{*}{pcre2}   & \FuDriver{}   & 14852 & 21273 & 22  \\
            & \FUDGE{} & 9144 & 13020  & 21   \\
        \midrule
\multirow{2}{*}{c\_ares} & \FuDriver{}   & 31    & 67    & 2       \\
        & \FUDGE{} & --     & --     & --          \\
           \midrule
\multirow{3}{*}{total} & \FuDriver{}   & 28478    & 61126    & 32      \\
        & \FUDGE{} &  13712    &    19973  &    22       \\    
        & increase & 1.08X$\uparrow$ & 2.06X$\uparrow$ & 0.45X$\uparrow$ \\
\end{mytable3}

% 待补充 >>>>>>>>>>>>>>>>>>>>>>>>>>>>

% \newcolumntype{C}{>{\arraybackslash}p{1.0cm}}
\NewEnviron{mytable_path1}[1]{
  \begin{table}[!htbp]
    \caption{#1}
    \scriptsize
    \scalebox{1}[1]{%
      \label{table-Path-FUDGE}
      \begin{tabular}{p{0.9cm}p{1.0cm}|CCCC}
        \toprule
        {\mysize Project}
        & {\mysize }
        & {\mysize driver1}
        & {\mysize driver2}
        & {\mysize driver3}
        & {\mysize total}\\
        \midrule
        \BODY
        \bottomrule
      \end{tabular}
    }
  \end{table}%
}{}

\begin{comment}
\begin{mytable_path1}{Number of paths (\FuDriver{}/FUDGE)}
\multirow{2}{*}{re2}     & FuDriver   & 11228  & 122  & 114   & 11464  \\
        & FUDGE  & 100  & 88   & 76    & 264  \\
        \midrule
\multirow{2}{*}{harfbuzz}& FuDriver   & 12708  & 3795  & 66  & 16569  \\
        & FUDGE & 3952 & 68    & 67        & 4087   \\
        \midrule
\multirow{2}{*}{guetzli} & FuDriver   & 7728   & 3016    & 238 & 10892   \\
        & FUDGE & 1460    & 123  & 113       & 1696    \\
        \midrule
\multirow{2}{*}{libjpeg} & FuDriver   & 780  & 762  & 59  & 1601  \\
        & FUDGE & 785  & 760    & 738  & 2283  \\
        \midrule
\multirow{2}{*}{woff2}   & FuDriver   & 3936    & 27    & 25    & 3988   \\
        & FUDGE & 59  & 51     & 23        & 133    \\
        \midrule
\multirow{2}{*}{json}    & FuDriver   & 3278   & 2078    & 1351    & 6707   \\
        & FUDGE & 632   & 196   & 118   & 946   \\
        \midrule
\multirow{2}{*}{libxml}  & FuDriver   & 128  & 25  & 3   & 156  \\
        & FUDGE & 12  & 3  & 3     & 18  \\
        \midrule
\multirow{2}{*}{pcre2}   & FuDriver   & 21273 & 14133 & 9513  & 44919 \\
            & FUDGE & 13020  & 12913  & 11530  & 37463 \\
        \midrule
\multirow{2}{*}{c\_ares} & FuDriver   & 67    & 2    & 2    & 71    \\
        & FUDGE & --     & --     & --     & --      \\
\end{mytable_path1}
\end{comment}

% 分析数据，结论：FuDriver的入口定位比fudge精准很多！ 
% TODO : 补充path: path更高+path可以直接求和
%\textbf{(1) Code coverage.} 
The third and fourth column of Table \ref{table-Branch-FUDGE} 
%and \ref{table-Path-FUDGE} 
presents the block coverage and path coverage statistics for each project. 
To obtain the total block coverage, we re-run each seed and merge all the blocks together. 
To obtain the total path coverage, we add the number of paths of each driver, since different drivers with different entry functions do not share the same path. 

We can observe that for each project, \FuDriver{} covers significantly more blocks than \FUDGE{}.
\FuDriver{} also covers much more paths than \FUDGE{} on most projects except \textit{libjpeg}, since we skip some similar high-level entry functions. 
Overall, \FuDriver{}'s block coverage outperforms that of \FUDGE{} by 79.54\%, while \FuDriver{} covers 105.52\% more paths than \FUDGE{}.
On projects such as \textit{harfbuzz}, \textit{guetzli}, \textit{woff2} and \textit{json}, \FuDriver{} is able to cover at least 100\% more blocks and paths than \FUDGE{}.
This is because \FuDriver{} tends to find high-level functions as potential entry functions, which usually call a series of low-level functions. However, \FUDGE{} only finds functions which call another function with the signature \textit{(const uint8\_t*, size\_t)}. These functions may be low-level functions and thus they are unable to cover much of the code.
The last two rows of Table~\ref{table-Branch-FUDGE} 
%and ~\ref{table-Path-FUDGE} 
show that on project \textit{c-ares}, \FUDGE{} fails to synthesize any valid fuzz drivers. Therefore no function in \textit{c-ares} calls another function with the target signature. However, \FuDriver{} can directly synthesize parameters for the entry function, which makes \FuDriver{} more versatile than \FUDGE{}.

The fifth column shows the unique bugs each driver detects. 
The fuzz drivers synthesized by \FuDriver{} trigger bugs in seven projects, while those synthesized by \FUDGE{} trigger bugs in only two projects. 
In total, \FuDriver{} finds 32 potential bugs, but \FUDGE{} only finds 22 potential bugs on \textit{json} and \textit{pcre2}, since \FuDriver{} tends to use high-level and more vulnerable functions as its entry functions, thus it is more likely to trigger potential bugs. 

% 总的结论
Based on the results shown in Table \ref{table-Branch-FUDGE},  
%, \ref{table-Path-FUDGE} 
%and \ref{table-Bug-FUDGE}, 
we can conclude that \FuDriver{} locates more effective entry functions, achieving higher block and path coverage and more unique bugs detected than \FUDGE{}, without the need for manual selection or modification.

% \begin{mdframed}
% \textit{Solution:}
% \textit{Collect precise feedback by inter-binary coverage linkage and bijective block mapping.}
% \end{mdframed}

\section{Case study on Real Projects}
\label{sec:Studies}
In this section, we use many real-world projects from Google's \textit{fuzzer-test-suite} and our industrial collaborators to demonstrate the effectiveness of \FuDriver{}.

\subsubsection{Real-world Project in Google's \textit{fuzzer-test-suite}}

%In this section, 
We first dissect the manually written driver and the fuzz drivers generated by \FuDriver{} and \FUDGE{} for the \textit{pcre2} library.

% original
\textbf{(1) The original manually written driver.} 
As presented in Listing \ref{list-4}, the driver written by domain experts invokes three entry functions: \textit{regcomp()}, \textit{regexec()} and \textit{regfree()}.  
First, it calls \textit{regcomp()} on the buffer generated by the fuzz engine.
Then, it calls \textit{regexec()} with the previous variable \textit{preg}. 
Finally, it calls \textit{regfree()} to free the variable \textit{preg}. 
%As we can see, the original driver is robust with a complete workflow.

\begin{figure}[!htbp]
\begin{lstlisting}[basicstyle=\footnotesize,caption={Driver written by domain experts},label=list-4]
extern "C" int LLVMFuzzerTestOneInput(const unsigned char *data, size_t size) 
{
    if (size < 1) return 0;
    regex_t preg;
    string str(reinterpret_cast<const unsigned char*>(data), size);
    string pat(str);
    int flags = data[size/2] - 'a';
    if (0 == regcomp(&preg, pat.c_str(), flags)) 
    {
        regmatch_t pm[5];
        regexec(&preg, str.c_str(), 5, pm, 0);
        regfree(&preg);
    }
    return 0;
}
\end{lstlisting}
\end{figure}

% FuDriver
\textbf{(2) The driver synthesized by \FuDriver{}.} 
\FuDriver{} locates \textit{pcre2\_match()} as one of the functions with the highest priority and regards it as the entry function for driver synthesis, whose function prototype is shown in Listing \ref{list-5}.

\begin{figure}[htb]
\begin{lstlisting}[basicstyle=\footnotesize,caption={Function pcre2\_match},label=list-5]
PCRE2_EXP_DEFN int PCRE2_CALL_CONVENTION pcre2_match(const pcre2_code *code, PCRE2_SPTR subject, PCRE2_SIZE length, PCRE2_SIZE start_offset, uint32_t options, pcre2_match_data *match_data, pcre2_match_context *mcontext);
\end{lstlisting}
\end{figure}

To synthesize fuzz drivers for the entry function \textit{pcre2\_match()}, \FuDriver{} needs to synthesize its parameters. 
First, \FuDriver{} regards the second and third parameters as the buffer and its size, then binds them with the buffer generated by the fuzz engine.
For the three complex parameters \textit{code}, \textit{match\_data} and \textit{mcontext} in \textit{pcre2\_match()}, \FuDriver{} calls function \textit{pcre2\_compile()} to get the variable \textit{code}, calls \textit{pcre2\_match\_data\_create\_from\_pattern()} to get the variable \textit{match\_data}. 
As for \textit{mcontext}, \FuDriver{} does not find a proper function to initialize it, so it synthesizes a variable with the same type and assigns all bits with 0.
In reality, the variable is a \textit{NULL} pointer.
As a result, \FuDriver{} synthesizes a candidate fuzz driver for function \textit{pcre2\_match()}.
We run the driver automatically and find that it results in a memory leak. 
To fix the memory leak, \FuDriver{} calls function \textit{pcre2\_match\_data\_free()} to free the variable \textit{match\_data}, and calls \textit{pcre2\_code\_free()} to free the variable \textit{code}.

Finally, \FuDriver{} synthesizes a valid fuzz driver with the two free functions, as presented in Listing~\ref{list-7}.
The driver first calls \textit{pcre2\_compile()} to obtain a pointer of type \textit{pcre2\_code()}.
Then it checks whether the \texttt{pointer} is \textit{NULL}.
If not, \textit{pcre2\_match\_data\_create\_from\_pattern()} is called to get a pointer of type \textit{pcre2\_match\_data()}. 
The driver also checks whether the acquired pointer is \textit{NULL}.
If not, the driver calls the entry function \textit{pcre2\_match()} with the generated arguments to initiate fuzzing.
Finally, it calls \textit{pcre2\_match\_data\_free()} and \textit{pcre2\_code\_free()} to free the two pointers previously generated.

\begin{figure}[!htb]
\begin{lstlisting}[basicstyle=\footnotesize,caption={Driver synthesized by \FuDriver{}},label=list-7]
uint32_t p1, p2, p5;
uint64_t p3, p4;
pcre2_code* v1 = pcre2_compile(data, size, p1, &p2, &p3, NULL);
if (v1)
{
    pcre2_match_data *v2 = 
        pcre2_match_data_create_from_pattern(v1, NULL);
    if (v2)
    {
        pcre2_match(v1, data, size, p4, p5, v2, NULL);
    }
    pcre2_match_data_free(v2);
}
pcre2_code_free(v1);
\end{lstlisting}
\end{figure}

\vspace{0.1cm}

\textbf{(3) The driver synthesized by FUDGE.} 
We also utilize \FUDGE{} to synthesize fuzz drivers. \FUZZGEN{} can not generate valid driver without providing additional test cases. 
First, \FUDGE{} finds a function \textit{compile\_pattern()} which calls another function \textit{pcre2\_compile()} with the function signature \textit{(const uint8\_t*, size\_t)}.
Then, \FUDGE{} extracts \textit{pcre2\_compile()}'s relevant code snippets in \textit{compile\_pattern()} and generates a driver.
Since \FUDGE{} does not propose a method to assign value for pointer parameters, 
we need to manually assign all pointer parameters to \textit{NULL}. 
In addition, this driver will cause a memory leak similar to the one found in \FuDriver{}. 
Since \FUDGE{} does not propose a way to free allocated memory correctly, we have to call \textit{pcre2\_code\_free()} manually. The final modified driver is shown in Listing~\ref{list-FUDGE-3}.

\begin{figure}[htb]
\begin{lstlisting}[basicstyle=\footnotesize,caption={Driver synthesized by FUDGE with modification},label=list-FUDGE-3]
extern "C" int LLVMFuzzerTestOneInput(const uint8_t* data, size_t size) {
    unsigned char buffer[8292] = {};
    memcpy(buffer, data, size);
    PCRE2_SIZE erroffset;
    int errcode;
    pcre2_compile_context 
        *compile_context = NULL;
    int options = data[0];
    pcre2_code * v1 = NULL;
    if (v1 != NULL) 
        return 0;
    v1 = pcre2_compile(buffer, -1, options, &errcode, &erroffset, compile_context);
    pcre2_code_free(v1);
    return 0;
}
\end{lstlisting}
\end{figure}

%\vspace{0.2cm}

\begin{figure}[!htb]
    \centering
    \subfigtopskip=2pt %设置子图与上面正文或别的内容的距离
    \subfigbottomskip=2pt %设置第二行子图与第一行子图的距离，即下面的头与上面的脚的距离
    \includegraphics[width=0.5\textwidth]{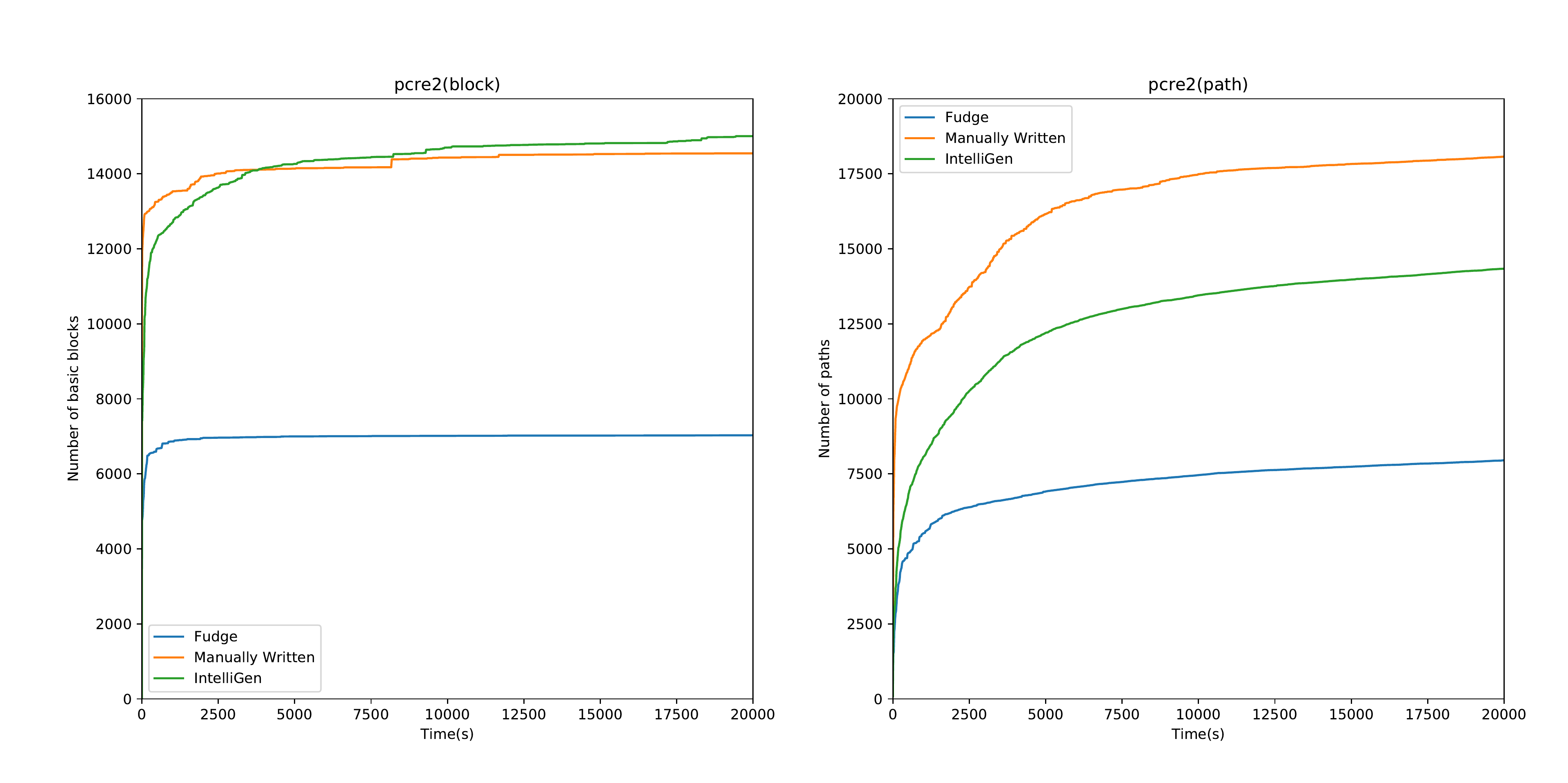}
    %\vspace{-0.5 cm}
    \caption{Block and path coverage for drivers written manually, and drivers synthesized by \FUDGE{} and \FuDriver{}.}
    \label{fig:new-pcre2}
\end{figure}

\textbf{(4) Code coverage of different drivers.} 
We run each driver for 24 hours, repeat 10 times and collect their average code coverage information as presented in Figure~\ref{fig:new-pcre2}.

% 走势

% 补充path相关
We can observe that the block coverage grows fast initially for all drivers, 
then the growth changes to an approximate logarithmic rate.
This is because the fuzzer tends to discover `easy' blocks rapidly, but once they have been mostly discovered, the fuzzer will require relatively more effort to cover the 'hard' blocks.
Also, there is a theoretical upper limit for the coverage of each driver: an inefficient driver that calls a low-level API function can not cover many blocks, while an efficient driver that calls a high-level API function can cover a lot of blocks after running a significant amount of time.
The path coverage trend is similar to that of the block coverage but continues to grow even after fuzzing for 24 hours. This is because while the seeds may cover the same set of blocks, the order in which they are visited constitute different paths.

We can also see that \FuDriver{} and \FUDGE{} both cover fewer blocks than the original driver written by domain experts in the first few seconds.
But after fuzzing for an substantial period, the driver synthesized by \FuDriver{} covers an increasing amount of blocks and can even cover a few more blocks than the manually written driver at the end of 24 hours. 
The driver synthesized by FUDGE, however, does not exhibit an obvious coverage growth after fuzzing for 24 hours.
This is because the driver synthesized by FUDGE does not contain the core function \textit{pcre2\_match()}.
If we manually modify the driver to call \textit{pcre2\_match()} or \textit{regexec()}, then it can cover more blocks after a substantial fuzzing period.

\subsubsection{Real-world projects from industrial collaborator}

We use \FuDriver{} on three real-world projects used by Huawei Technologies: \textit{mxml v2.9}, \textit{mxml v2.12} and \textit{libevent}. These three projects are also widely used in other industrial communities. The overall results are presented in Table~\ref{table-mxml-libevent-fudriver}, where \FUZZGEN{} did not generate valid drivers because of the absence of test cases. The first, second and third column represents the number of blocks covered in each project, the number of paths covered and the number of unique bugs detected, respectively.

% 回来改这个table

% >>>>>>>>>>>>>>>>>>>>>>>>>>>>>>>>>>>>

% ==========================================

% \newcolumntype{C}{>{\arraybackslash}p{1.5cm}}
\NewEnviron{FuDriver_practice}[1]{
  \begin{table}[!htb]
    \caption{#1}
    \scriptsize
    \scalebox{1}[1]{%
      \label{table-mxml-libevent-fudriver}
      \begin{tabular}{p{0.9cm}p{1.0cm}|CCCC}
        \toprule
        {\mysize Project}
        & {\mysize }
        & {\mysize Blocks}
        & {\mysize Paths}
        & {\mysize Unique bugs}\\
        \midrule
        \BODY
        \bottomrule
      \end{tabular}
    }
  \end{table}%
}{}

\begin{FuDriver_practice}{Number of blocks, paths, and unique bugs}
\multirow{2}{*}{mxml v2.9}   & \FuDriver{}   & 715  & 503  & 5   \\
        & FUDGE & 31   & 43  & 0  \\
        \midrule
\multirow{2}{*}{mxml v2.12} & \FuDriver{}   & 735  & 502  & 5   \\
        & FUDGE & 31  & 42    & 0   \\
        \midrule
\multirow{2}{*}{libevent} & \FuDriver{}   & 414   & 11    & 0    \\
        & FUDGE & 113    & 3  & 0    \\
\end{FuDriver_practice}

% <<<<<<<<<<<<<<<<<<<<<<<<<<<<<<<<<<<<<<<<<<<<<<<<<

%According to the table, 
The statistics show that \FuDriver{} covers more blocks and paths and can detect more bugs than \FUDGE{} in all cases tested. This is because \FuDriver{} tends to synthesize the fuzz drivers on high-level entry functions while \FUDGE{}'s synthesis criteria is limited to a specific function signature.
In addition, \FuDriver{} and FUDGE cover more blocks and paths on \textit{mxml v2.9} and \textit{mxml v2.12} than on \textit{libevent}. %, but actually \textbf{libevent} is a much larger project than \textbf{mxml}.
This is because \textbf{libevent} is stateful, thus requiring the driver to call a series of functions in a particular order before starting the whole project. One way to improve this is to let \FuDriver{} learn the order in which to call multiple entry functions from unit test cases. %, which is one of our future work.

\section{Lesson Learned}
\label{sec:LessonLearned}
% We now present the 
% This section describes what we learned from the development and implementation of \FuDriver{}.

% 手写一个Driver很困难->为什么困难(花费了大量的时间在读源代码并挑选合适的入口)。->工业界对Fuzz的需求很广泛，手写Driver很费时->Driver生成可以有效的解决这个问题。
% 和Evaluation的Case Study联系一下，FUDGE哪里不好，为什么手写Driver很耗时

% Lesson 1
% 【这地方不是重点，可以简单的写】Fuzz在工业实践中是一个很好的漏洞挖掘手段，目前也有各式各样的Fuzz工具在不断的推出并挖掘出许多漏洞。
%【重点1】然而目前<<限制>>工业FUzz发展的一个重要原因是<<需要人工进行Fuzz驱动的适配adaption>>[突出这点！！！]。通过自动化的手段能够大大减少适配的难度。
%【重点2】特别是对于大型的应用来说，往往一个驱动是不够的，然而人工适配驱动代价又很大。通过我们的方法可以快速定位多个驱动，实验证明驱动效果差不多

% 选择正确的入口函数很重要，这地方可能
% 正确合成参数很重要，lazy-store操作是必须的
% 有限的时间内，fuzz更少的函数可能效果更好
% 自动生成驱动尽量不要依赖于测试用例

%\textbf{Pass precise arguments to the entry functions is very important for fuzz testing.} To call the entry function, we need to 
%\textbf{}
% ----------------------------新加完

\textbf{Writing a fuzz driver manually seriously hinders the efficiency of fuzz testing.} Fuzzing tools have been well developed and widely deployed in industrial environments and have detected many vulnerabilities. 
However, the overall performance and effectiveness of fuzz testing is still below expectations, since testers need to undertake the extremely laborious task of constructing fuzz drivers manually for each project.
This renders fuzz testing inaccessible to many who wish to use fuzz testing. 
Generating fuzz drivers automatically can greatly reduce the amount of manual labor required, especially for large projects, which usually demands a diverse portfolio of fuzz drivers to cover most areas of the program's code.
%But the human resources consumption on writing multiple fuzz drivers is very high.

%Our experiment also proves this conclusion.
%From column ``Original'' and ``FuDriver-Total'' of table \ref{table-1}, the results show that \FuDriver{} can almost always cover more branches than original driver, if \FuDriver{} synthesizes enough fuzz drivers.

% Lesson 2
%【重点】驱动的质量对Fuzz影响很大->（详细点，为什么影响很大（high low level，highlevel有时有语法检查等，需要扩充））->实验证明影响很大（我们表X也说明了。。。）（同一个项目，相同的算法生成的驱动，不同的入口，质量结果差很多。相同入口，FUDriver和人工的差不多）【不要提FUDGE！！！】->

\textbf{The quality of fuzz drivers will drastically impact the performance of fuzz testing.}
A fuzz driver should select a high-value entry function to maximize its effectiveness. 
Different entry functions can reach different parts of the code and will determine the direction of the fuzzing. 
A high-level function usually calls many low-level functions, thus calling a high-level function usually covers more branches and paths than that of calling a low-level function.
However, sometimes a high-level function may contain numerous error checking code, making it difficult for the fuzzer to reach low-level code.
An efficient fuzz driver should attempt to bypass these error checking code and reach the core of a project directly.
The experiment results show that the choice of entry functions has a great influence on fuzzing efficiency, and we need more intelligent methods to decide the entry function. 

\textbf{The performance of driver synthesis can be improved with more domain knowledge.}

% Currently, \FuDriver{} can only synthesize fuzz drivers for only one entry function.

As we have explained previously, some projects are stateful and will require the fuzz driver to call a series of functions in a specific order to reach most of the code. 
In addition, our argument construction algorithm may produce semantically incorrect entry function parameters. Through leveraging domain knowledge automatically extracted from test cases and other programs that use this library, \FuDriver{} can solve the aforementioned problems through learning an entire sequence of function calls and understanding which function parameters would be accepted by the library, respectively.

\textbf{The criteria for identifying effective entry functions is largely undetermined.} Though \FuDriver{} utilizes the \EFL{} to locate potential entry functions using metrics that represent potential memory vulnerabilities, to avoid generating sub-optimal fuzz drivers, we still manually select the entry functions that potentially allow the fuzz engine to cover a large amount of the code. A concrete selection criteria is needed to further filter out less potential entry functions and maximize the effectiveness of generated fuzz drivers. Nevertheless, our evaluation is valid regardless of this procedure, as a fully automated procedure would simply generate more fuzz drivers. 

% \section{Discussion}
% \label{Discussion}
%   \input{part-discussion}

% \section{future work}
% \input{part-future}

\section{Conclusion}
\label{sec:Conclusion}
In this paper, we propose \FuDriver{}, an automated fuzz driver synthesis framework. It constructs valid fuzz drivers using the following steps. \FuDriver{} first calculates the vulnerability priority for each function and takes the functions with the highest priority as entry functions. Then, \FuDriver{} mutates the arguments of functions called in an entry function and synthesizes parameters for the entry function using algorithmic parameter construction to synthesize a fuzz driver. We evaluate \FuDriver{}'s effectiveness against state-of-the-art fuzz driver synthesizers \FUZZGEN{} and \FUDGE{} on real-world projects selected from the Android Open Source Project, Google's \textit{fuzzer-test-suite} and our industrial collaborators. Compared with \FUZZGEN{}, \FuDriver{} covers 2.03$\times$ more blocks and 1.36$\times$ more paths. Compared with \FUDGE{}, \FuDriver{} covers 1.08$\times$  more blocks, 2.06$\times$ more paths, and detects ten more bugs. 
\FuDriver{} constitutes a significant improvement over the current state-of-the-art, providing vastly improved driver synthesis quality with a vast reduction in manual intervention, making fuzz testing more accessible and versatile.

% 学习聪哥 提前到通讯作者那个地方
\begin{comment}
\section*{Acknowledgement}
This research is sponsored in part by the NSFC Program (No. 62022046, U1911401, 61802223), National Key Research and Development Project (Grant No. 2019YFB1706200), the Huawei-Tsinghua Trustworthy Research Project (No. 20192000794).
\end{comment}

\begin{comment}
Our future work mainly focus on two aspects. First, we can utilize \FuDriver{} to analyze the \textit{main} function or test cases of a library to learn how we should call an entry function with appropriate parameters, and how to call multiple entry functions in a correct order. %Because test cases or the main function usually contain a complete calling process of the target project.
Second, \FuDriver{} can learn how to pass correct parameters to a low-level entry function from the annotation or documentation of the target project, which can help \FuDriver{} avoid API misuses. 
%Third, we can make \FuDriver{} track each fuzz driver's coverage, to choose the more efficient fuzz drivers automatically.

% In this papaer, to solve XXX problem， we propse \FuDriver{}, which XXXX, XXX. 然后在XXXevalution， 效果怎么样，
\end{comment}

\normalsize \bibliographystyle{acm}
\bibliography{ADG}

\end{document}